\documentclass[aps,prl,floatfix,amsmath,amssymb,superscriptaddress,showpacs,showkeys,onecolumn,8pt]{revtex4-2}
\usepackage[caption=false]{subfig}
\usepackage{graphicx,bm,color}
\usepackage{amsfonts}
\usepackage{color}
\usepackage{lipsum, color}
\usepackage{amsthm}

\usepackage{braket}
\newtheorem{theorem}{Theorem}
\usepackage{hyperref}
\hypersetup{
    colorlinks=true,
    linkcolor=blue,
    filecolor=blue,      
    urlcolor=blue,
}
\usepackage{epstopdf}
\begin{document}

\frenchspacing
\title{Maximum Violation of Monogamy of Entanglement for Indistinguishable Particles by Measures that are Monogamous for Distinguishable Particles}
\author{Goutam Paul}
\email{goutam.paul@isical.ac.in}
\affiliation{Cryptology and Security Research Unit, R. C. Bose Centre for Cryptology and Security, Indian Statistical Institute, Kolkata 700108, India}
\author{Soumya Das}
\email{soumya06.das@gmail.com}
\affiliation{Cryptology and Security Research Unit, R. C. Bose Centre for Cryptology and Security, Indian Statistical Institute, Kolkata 700108, India}
\author{Anindya Banerji}
\email{abanerji09@gmail.com}
\affiliation{Quantum Science and Technology Laboratory, Physical Research Laboratory, Ahmedabad 380009, India}

\begin{abstract}
Two important results of quantum physics are the \textit{no-cloning} theorem and the \textit{monogamy of entanglement}. The former forbids the creation of an independent and identical copy of an arbitrary unknown quantum state and the latter restricts the shareability of quantum entanglement among multiple quantum systems. For distinguishable particles, one of these results imply the other. In this Letter, we show that in qubit systems with indistinguishable particles (where each particle cannot be addressed individually), a maximum violation of the monogamy of entanglement is possible by the measures that are monogamous for distinguishable particles. To derive this result, we formulate the degree of freedom trace-out rule for indistinguishable particles corresponding to a spatial location where each degree of freedom might be entangled with the other degrees of freedom. Our result removes the restriction on the shareability of quantum entanglement for indistinguishable particles, without contradicting the no-cloning theorem.
\end{abstract}


\maketitle
\emph{Introduction.}---  
An interesting feature of quantum physics is the presence of {\em identical particles} and their distinguishability. Throughout this Letter,  by identical particles~\cite{Feynman94,Sakurai94} we mean a set of particles with the same physical properties, except possibly their spatial locations; and by {\em indistinguishable particles}~\cite{Ghirardhi02}, we mean a set of identical  particles that cannot be labeled separately even by their spatial locations. Such particles find applications in Bose-Einstein condensate~\cite{Morsch06,Esteve08}, quantum metrology~\cite{Giovannetti06,Benatti11}, quantum dots~\cite{Petta05,Tan15}, ultracold atomic gases~\cite{Leibfried03} and as a resource~\cite{Morris20} for tasks like teleportation~\cite{Ugo15,LFC18}, entanglement swapping~\cite{LFCES19,Das20} etc. 

\begin{figure}[htbp] 
\centering
\includegraphics[width=8.6cm]{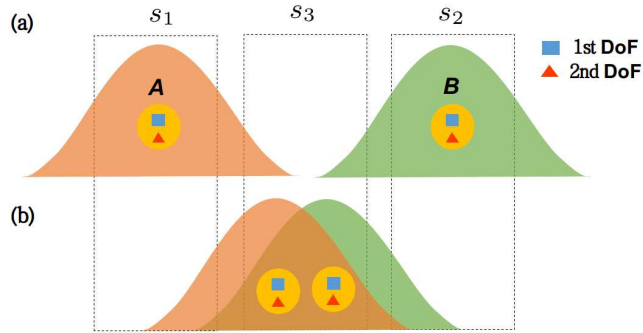} 
\caption{Creation of indistinguishable particles. (a) Two identical particles $A$ and $B$ with two degrees of freedom (DoFs) each, denoted by square and triangle shapes, are present in distinct spatial locations $s_1$ and $s_2$ in such a way that their wave-functions do not overlap. Though identical, they are distinguishable via their spatial locations. (b) The particles are brought close to each other so that their wave-functions overlap and they become indistinguishable. If the measurement is done in any DoF in the overlapped region, i.e., $s_3$, then it is not possible to detect which particle is measured. Even if the particles are again moved apart, they can no longer be labeled. The information about which of $A$ and $B$ appears at $s_1$ or $s_2$ is lost.}
\label{Wave_overlap}
\end{figure}

One method of producing indistinguishable particles  starting from identical ones is known as particle exchange \cite{Y&S92PRA,Y&S92PRL}. Figure~\ref{Wave_overlap} shows a conceptual representation of this process.

 \begin{figure*}[t!] 
\centering
\includegraphics[width=17.2cm]{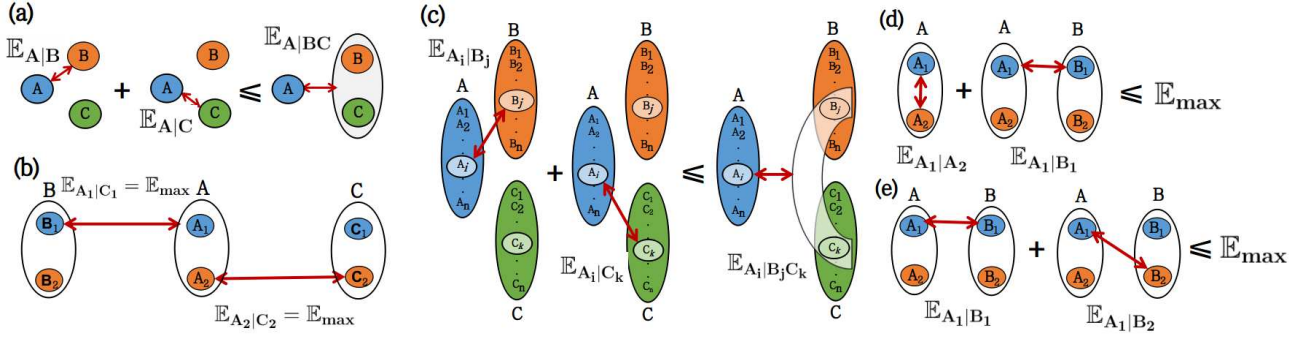} 
\caption{Consider three particles $A$, $B$, $C$, and a bipartite entanglement measure $\mathbb{E}$ where $\mathbb{E}_{X|Y}$ measures the entanglement between the subsystems $X$ and $Y$ of the composite system $XY$ and $\mathbb{E}_{max}$ denotes its maximum value. Now consider the following five scenarios: (a) Particle-based MoE obeying Eq.~\eqref{Gen_Monogamy}. (b) Here, $A$ is maximally entangled with $B$ in  DoF 1 (i.e., $\mathbb{E}_{A_{1} \mid B_{1}}=\mathbb{E}_{max} $) and with $C$ in DoF 2 (i.e., $\mathbb{E}_{A_{2} \mid C_{2}}=\mathbb{E}_{max} $). In particle view, apparently MoE is violated; but in DoF view, it is not. (c) Inter-DoF MoE proposed in Eq.~\eqref{Degree_Monogamy} which resolves the previous apparent violation. (d) \& (e) Two-particle inter-DoF MoE, where $\mathbb{E}_{A_{1} \mid A_{2}}$ measures entanglement between the two DoFs $A_{1}$, $A_{2}$ of $A$ and $\mathbb{E}_{A_{1} \mid B_{1}}$  between $A_{1}$, $B_{1}$ (in (d)); and $\mathbb{E}_{A_{1} \mid B_{j}}$ between $A_{1}$ of $A$ and $B_{j}$ of $B$, $j \in \lbrace 1, 2\rbrace$ (in (e)).}
\label{Mono}
\end{figure*}

A composite system of indistinguishable particles cannot be decomposed in terms of distinguishable subsystems, as the underlying Hilbert space structure is fundamentally different~\cite{Ghirardhi02}. This difference becomes crucial when one considers entanglement of distinguishable vs. indistinguishable particles~\cite{Li01,You01,John01,Zanardi02,Wiseman03,Ghirardi04,Vedral03,Barnum04,Barnum05,Zanardi04,Benatti14,Benatti20}. While the notion of entanglement for distinguishable particles corresponds to the tensor product of the corresponding Hilbert spaces~\cite{HHHH08,Nielsenbook}, for indistinguishable case one must consider either the symmetric (bosonic) or the antisymmetric (fermionic) subspace of the whole Hilbert space~\cite[Sec. III]{Barun18}. This leads to either particle-based first quantization~\cite{Li01,You01,John01,Zanardi02,Ghirardhi02,Wiseman03,Ghirardi04} or mode-based second quantization~\cite{Vedral03,Barnum04,Barnum05,Zanardi04} approaches giving contradictory results by representing some uncorrelated states as entangled ones~\cite{Ghirardi04,Tichy11}. To settle this issue, a new approach was proposed in~\cite{LFC16,LFC18} which we use as a definition of entanglement of indistinguishable particles throughout this Letter (see Supplemental Material~\cite[Appendix A]{Supply}). 

One important feature of quantum entanglement of distinguishable particles~\cite{HHHH08} is its restriction upon the shareability among composite systems (consisting of particles or degrees of freedom (DoFs)), known as~\emph{monogamy of entanglement} (MoE). It has applications in key distribution~\cite{Pawlowski10}, quantum games~\cite{Tomamichel13,Johnston16}, state classification~\cite{Dur00},  
 etc.~\cite{Bae06,Chiribella06,Lloyd14,Ma11,Garcia13,Brandao15,Rao13}.

 A bipartite entanglement measure $\mathbb{E}$ that obeys the relation 
\begin{equation} \label{Gen_Monogamy}
\mathbb{E}_{A|B}(\rho_{AB})+\mathbb{E}_{A|C}(\rho_{AC}) \leq \mathbb{E}_{A|BC}(\rho_{ABC}),
\end{equation} 
for all $\rho_{ABC}$ where $\rho_{AB}=\text{Tr}_{C} \left( \rho_{ABC}\right) $, $\rho_{AC}=\text{Tr}_{B} \left( \rho_{ABC}\right) $, $\mathbb{E}_{X|Y}$ measures the entanglement between the systems $X$ and $Y$ of the composite system $XY$, and the vertical bar represents bipartite splitting, is called \textit{monogamous}. Such inequality was first shown for squared concurrence ($\mathcal{C}$)~\cite{Hill97,Wootters98} by Coffman, Kundu and Wootters (CKW) for three parties~\cite{CKW00} and later generalized for $n$ parties~\cite{Osborne06}.

Some commonly used monogamous entanglement measures for qubit systems are the entanglement of formation~\cite{Bennett96,Oliveira14,Bai14PRL,Bai14PRA,Gou19}, log-negativity~\cite{Zyczkowski98,Vidal02}, Tsallis-q entropy~\cite{Kim10,Luo16}, R\'{e}nyi-$\alpha$ entanglement~\cite{Kim_Sanders10,Song16}, Unified-(q, s) entropy~\cite{Kim11,Khan19}, etc.~\cite{Plenio07,Guine09}. For higher dimensional systems, squared concurrence is known to violate~\cite{Ow07} Eq.~\eqref{Gen_Monogamy}, and only a few entanglement measures are monogamous like one-way distillable entanglement~\cite{Devetak05} and squashed entanglement~\cite{Christandl04,Brandao11}. Throughout this Letter, we focus on MoE and its violation using bipartite entanglement measures in qubit systems only. 

Suppose a bipartite entanglement measure $\mathbb{E}$ attains the maximum value $\mathbb{E}_{max}$ for maximally entangled states. Consider a situation when $\mathbb{E}_{A|B}(\rho_{AB}) < \mathbb{E}_{max}$, $\mathbb{E}_{A|C}(\rho_{AC}) < \mathbb{E}_{max}$, but  $\mathbb{E}_{A|B}(\rho_{AB}) + \mathbb{E}_{A|C}(\rho_{AC}) >\mathbb{E}_{max}$. Obviously, this causes a violation of MoE which we call a \textit{non-maximal violation}. Consider another situation, when 
\begin{equation} \label{MaxVio}
\mathbb{E}_{A|B}(\rho_{AB}) = \mathbb{E}_{max} \mbox{ and } \mathbb{E}_{A|C}(\rho_{AC}) = \mathbb{E}_{max}, 
\end{equation}
i.e., when $A$ is maximally entangled with both $B$ and $C$, we call the corresponding violation as the \textit{maximal violation} of MoE. 
For qubit systems with distinguishable particles, the first situation above would not lead to a violation of the no-cloning theorem~\cite{QC05,QC14}, but the second situation would do (see Supplemental Material~\cite[Appendix B]{Supply}).

The above result holds irrespective of whether $A$, $B$ and $C$ are single-DoF particles or DoFs of the same/different particles as shown in Fig.~\ref{Mono} (a) and (b). The entanglement measures which are monogamous for distinguishable particles are also so for systems of indistinguishable particles, where $A$, $B$, and $C$ are distinct spatial locations~\cite{Vedral03,Wiseman03,Bose13} with one particle each. However, interesting scenarios might arise when the involved particles are entangled in multiple DoFs which we investigate here. 

In this Letter, we establish a generalized DoF trace-out rule that covers single or multiple DoF scenarios for both distinguishable and indistinguishable systems. \textit{Partial trace-out operation}~\cite{HHHH08,Nielsenbook} is a typical method of finding the reduced density matrix of a subsystem which can be either one whole particle or a single DoF for distinguishable systems. However, for indistinguishable systems, applying the above method results in a contradiction in identifying entanglement~\cite{Ghirardi04,Tichy11}. Experimental works on such systems~\cite{Bloch08,Hayes07,Anderlini07,Petta05,Lundskog14,Tan15,Veldhorst15} existed earlier, but a common mathematical framework for a consistent theoretical interpretation was first attempted in~\cite{LFC16,LFC18}, by providing a method of partial trace-out for a whole indistinguishable particle. One may be tempted to think that the same rule can trace out a single DoF also. However, this is not so straightforward. When particles become indistinguishable, performing the partial trace-out of a particular DoF is challenging, because a DoF cannot be associated with a specific particle. We propose a method to perform partial trace-out of a DoF when the particles are indistinguishable by suitably modifying the framework of~\cite{LFC16,LFC18}. Our generalized method covers in a unified manner DoF or particle trace-out for single or multiple DoF scenarios for both distinguishable and indistinguishable particles. Using this generalized DoF trace-out, we show that MoE can be violated maximally by indistinguishable particles in qubit systems for measures (such as squared concurrence, log-negativity, etc.) that are monogamous for distinguishable particles. This result establishes a new fundamental difference between distinguishable and indistinguishable systems. In the former case, the no-cloning theorem and non-sharability of maximal entanglement (i.e., MoE) are equivalent. However, in the latter case, the no-cloning theorem remains valid, but maximal entanglement can be shared violating MoE.

\emph{Inter-DoF MoE}.--- Here we reformulate Eq.~\eqref{Gen_Monogamy} in a more general framework to include multiple DoFs of the same/different particles/entities. Although this is not a contribution, we include it here to establish the background for subsequent analysis.

Consider three entities $A$, $B$, and $C$, each with $n$ DoFs, numbered 1 to $n$. If the joint state of the $i$-th, $j$-th and $k$-th DoFs of $A$, $B$, and $C$ respectively is represented by $\rho_{A_{i}B_{j}C_{k}}$, then the inter-DoF MoE can be formulated as
\begin{equation} \label{Degree_Monogamy}
\mathbb{E}_{A_{i}|B_{j}}(\rho_{A_{i}B_{j}})+\mathbb{E}_{A_{i}|C_{k}}(\rho_{A_{i}C_{k}}) \leq \mathbb{E}_{A_{i}|B_{j}C_{k}}(\rho_{A_{i}B_{j}C_{k}}),
\end{equation} 
where $\rho_{A_{i}B_{j}}=\text{Tr}_{C_{k}} ( \rho_{A_{i}B_{j}C_{k}}) $, $\rho_{A_{i}C_{k}}=\text{Tr}_{B_{j}} ( \rho_{A_{i}B_{j}C_{k}}) $, and $\mathbb{E}_{X_{i}|Y_{j}}$ measures the entanglement between subsystems $X_{i}$ and $Y_{j}$ of the composite system $X_{i}Y_{j}$ as is shown in Fig.~\ref{Mono}~(c). It means that if the $i$-th DoF of $A$ is maximally entangled with the $j$-th DoF of $B$, then it cannot share any correlation with the $k$-th DoF of $C$. 

The inter-DoF MoE of Eq.~\eqref{Degree_Monogamy} is more general than the particle-based MoE of Eq.~\eqref{Gen_Monogamy}. The former includes the latter when the three DoFs $i$, $j$, and $k$ belong to three different particles $A$, $B$, and $C$ respectively. However, the inter-DoF MoE can capture many other scenarios that are illustrated in Fig.~\ref{Mono}~(d) and (e). 
 
Two interesting types of MoE involving only two particles can also be explained using the inter-DoF formulation. 

(i) \textit{Type I}: Here, MoE is calculated using $\mathbb{E}_{A_{i} \mid A_{j}}$ and $\mathbb{E}_{A_{i} \mid B_{k}}$, as shown in Fig.~\ref{Mono}~(d). Equation~\eqref{Degree_Monogamy} can capture this scenario by setting $A=B$. The recent analysis for distinguishable particles in~\cite{camalet17,camalet18} is a specific example of this type.

(ii) \textit{Type II}: Here, MoE is calculated using $\mathbb{E}_{A_{i} \mid B_{j}}$ and $\mathbb{E}_{A_{i} \mid B_{k}}$, as shown in Fig.~\ref{Mono}~(e). Equation~\eqref{Degree_Monogamy} can capture this scenario by setting $B=C$.

This formulation also includes the case of single-particle entanglement~\cite{Zanardi02,Karimi10,Hasegawa03}, when all the three DoFs come from a single particle. Equation~\eqref{Degree_Monogamy} can capture this scenario by setting $A=B=C$. Further, inter-DoF MoE is also valid for indistinguishable particles where the labels $A$, $B$, and $C$ denote spatial locations with each mode containing exactly one particle and $i$, $j$, and $k$ represents the DoFs at each spatial mode.

\emph{DoF trace-out for indistinguishable particles and its physical significance.}---
We have already discussed that the trace-out operation of~\cite{LFC16,LFC18} for indistinguishable particles is not readily applicable to trace out DoFs of such particles, particularly when the particles are entangled in multiple DoFs. In order to treat the cases of both distinguishable and indistinguishable particles under a uniform mathematical framework, we define trace-out of DoFs, rather than that of whole particles, by suitably modifying the formulation of~\cite{LFC16,LFC18}. 

Assume two indistinguishable particles each having two DoFs are associated with spatial labels  $\alpha$ and $\beta$. The $i$-th and the $j$-th DoFs are represented by $a_{i}$ and $b_{j}$ respectively, where $i, j \in \mathcal{N} =\lbrace 1, 2 \rbrace$. The general state of such a system is written as 
\begin{equation} \label{IDstate}
\begin{aligned}
\ket{\Psi^{(2)}} = \sum_{\alpha, a_{1}, a_{2}, \beta,  b_{1}, b_{2}} \kappa^{\alpha a_{1} a_{2}}_{\beta  b_{1} b_{2}} \ket{\alpha a_{1} a_{2}, \beta  b_{1} b_{2}},
\end{aligned}
\end{equation}
where $\alpha, \beta$ ranges over $ \mathcal{S} = \left\lbrace s_{1}, s_{2}, \cdots , s_{p} \right\rbrace $ which refers to distinct spatial locations with $p \geq 2$. Each of  $a_{i}, b_{i}$ ranges over  $ \mathcal{D}_{i} = \left\lbrace D_{i_{1}}, D_{i_{2}}, \cdots , D_{i_{q_{i}}} \right\rbrace $ which refers to the eigenvalues of the $i $-th  DoF, where $q_{i} \geq 2$, since each DoF must have at least two distinct eigenvalues. 

The value of $q_{i}$ may vary with $i$. For example, consider two DoFs: polarization and optical orbital angular momentum (OAM), associated with a system of indistinguishable photons. Generally, the polarization belongs to a two-dimensional Hilbert space, whereas the OAM lies in an infinite-dimensional Hilbert space governed by the azimuthal index $l$. In practical implementations, this mismatch in Hilbert space dimensions between the two DoFs is taken care of by mapping the larger dimensional space to the lower dimensional one~\cite{Karimi10,Bhatti15}. For OAM, the infinite-dimensional Hilbert space is generally mapped into a two-dimensional one with the eigenvalues $ \lbrace 2l, 2l+1 \rbrace$ or $\lbrace +l, -l \rbrace$.  Also, the Hilbert space is sometimes restricted to smaller dimensions by proper state engineering in which case only certain chosen values of $l$ are allowed.

 The general density matrix is expressed as 
\frenchspacing
 \medmuskip=-1mu
\thinmuskip=-1mu
\thickmuskip=-1mu 
\begin{equation} \label{IDDM}
\begin{aligned}
\text{ \begin{small}
$\rho^{(2)}~ = \sum_{\substack{ \alpha, \beta, \gamma, \delta,  a_{1}, a_{2}, \\ b_{1}, b_{2},  c_{1}, c_{2}, d_{1}, d_{2}}} \kappa^{\alpha a_{1} a_{2}}_{\beta  b_{1} b_{2}} \kappa^{\gamma c_{1} c_{2} *}_{\delta  d_{1} d_{2}}  \ket{\alpha a_{1} a_{2}, \beta  b_{1} b_{2}} \bra{\gamma c_{1} c_{2}, \delta  d_{1} d_{2}}$,
\end{small} }
\end{aligned}
\end{equation}
\frenchspacing
 \medmuskip=2mu
\thinmuskip=2mu
\thickmuskip=2mu 

\noindent where $ \alpha, \beta ,\gamma , \delta$ span $ \mathcal{S} $ and $a_{i}, b_{i}, c_{i}, d_{i}$ span $ \mathcal{D}_{i}$.
To perform  DoF trace-out of the $i$-th DoF, $i \in \mathcal{N}$, of spatial region $s_x \in \mathcal{S}$, we define the reduced density matrix as

\frenchspacing
 \medmuskip=0mu
\thinmuskip=0mu
\thickmuskip=0mu 
\begin{equation} \label{DoFTrRule}
\begin{aligned}
\rho_{s_{x_{\bar{i}}}} \equiv  \text{Tr}_{s_{x_{i}}} \left(  \rho^{(2)} \right) \equiv \sum_{m_{i} \in \mathcal{D}_{i}} \braket{ s_x m_i \mid \rho^{(2)} \mid s_x m_i } :=& \sum_{m_{i}} \bigg \lbrace   \sum_{\substack{ \alpha, \beta, a_{i}, a_{\bar{i}},  b_{1}, b_{2}, \\ \gamma, \delta, c_{i}, c_{\bar{i}}, d_{1}, d_{2}} } \kappa^{\alpha a_{i} a_{\bar{i}}}_{\beta b_{1} b_{2}} \kappa^{\gamma c_{i} c_{\bar{i}} *}_{\delta d_{1} d_{2}}  \braket{s_x m_i \mid \alpha a_i} \braket{\gamma c_i  \mid s_x m_i}  \ket{\alpha a_{\bar{i}}, \beta  b_{1} b_{2}}\bra{\gamma c_{\bar{i}}, \delta  d_{1} d_{2}}  \\
& +  \eta \sum_{\substack{ \alpha, \beta,  a_{1}, a_{2}, b_{i}, b_{\bar{i}}, \\ \gamma, \delta, c_{i}, c_{\bar{i}}, d_{1}, d_{2}} } \kappa^{\alpha a_{1} a_{2}}_{\beta b_{i} b_{\bar{i}}} \kappa^{\gamma c_{i} c_{\bar{i}} *}_{\delta d_{1} d_{2}} \braket{s_x m_i \mid \beta b_i }  \braket{\gamma c_i  \mid s_x m_i}  \ket{\alpha a_{1} a_{2}, \beta  b_{\bar{i}}} \bra{\gamma c_{\bar{i}}, \delta  d_{1} d_{2}} \\
& + \eta  \sum_{\substack{ \alpha, \beta, a_{i}, a_{\bar{i}},  b_{1}, b_{2}, \\ \gamma, \delta, c_{1}, c_{2} , d_{i}, d_{\bar{i}}} } \kappa^{\alpha a_{i} a_{\bar{i}}}_{\beta b_{1} b_{2} } \kappa^{\gamma c_{1} c_{2} *}_{\delta d_{i} d_{\bar{i}}}  \braket{s_x m_i \mid \alpha a_i}  \braket{\delta d_i  \mid s_x m_i}  \ket{\alpha a_{\bar{i}}, \beta  b_{1} b_{2}}\bra{\gamma c_{1} c_{2}, \delta  d_{\bar{i}}}   \\
 & + \sum_{\substack{ \alpha, \beta,  a_{1}, a_{2}, b_{i}, b_{\bar{i}}, \\ \gamma, \delta, c_{1}, c_{2}, d_{i}, d_{\bar{i}}} } \kappa^{\alpha a_{1} a_{2}}_{\beta b_{i} b_{\bar{i}}} \kappa^{\gamma c_{1} c_{2} *}_{\delta d_{i} d_{\bar{i}}}  \braket{s_x m_i \mid \beta b_i } \braket{\delta d_i  \mid s_x m_i}   \ket{\alpha a_{1} a_{2}, \beta  b_{\bar{i}}} \bra{\gamma c_{1} c_{2}, \delta  d_{\bar{i}}} \bigg \rbrace,
\end{aligned}
\end{equation}
\frenchspacing
 \medmuskip=2mu
\thinmuskip=2mu
\thickmuskip=2mu 
\noindent where $\bar{i}:=(3-i)$. The parameter  $\eta$ is $+1$ ($-1$) for bosons (fermions). 
Equation~\eqref{DoFTrRule} can be generalized for $n$ DoFs and it includes particle trace-out as a special case for $n=1$ (see Supplemental Material~\cite[Appendix C]{Supply}). 
 
 Our DoF trace-out rule plays a very critical role with respect to the recently introduced complex systems with inter-DoF entanglements~\cite{HHNL,camalet17,camalet18}. When such entanglement exists, measuring or non-measuring one of the participating DoFs would influence the measurement results of the other participating DoFs. The statistics so obtained cannot be predicted using the existing trace-out rules in either the first or second quantization notations. Equation~\eqref{DoFTrRule}, on the other hand, can deal with all such systems with inter-DoF correlations in indistinguishable particles, leading to the prediction of perfect measurement statistics. Further, it generalizes the standard existing trace-out rule and is therefore suitable for such entanglement structures of distinguishable particles as well. More specifically, for distinguishable particles, tracing out a single DoF of a particle is analogous to tracing out a whole particle; for indistinguishable particles, on the other hand, tracing out a single DoF is performed for a specific spatial location where wave-functions of multiple particles might be overlapping. These overlaps are taken care of in the inner-product terms in the expression of Eq.~\eqref{DoFTrRule}. 

\emph{Violation of MoE by indistinguishable particles.---} 
The inter-DoF MoE is not absolute and can be violated maximally by indistinguishable particles. For illustration, consider two-particle inter-DoF entanglement~\cite[Eq. (4)]{HHNL} between spin and path. It can be represented as $\ket{\Psi^{(2)}}$ of Eq.~\eqref{IDstate} with the parameters $\alpha, \beta \in \lbrace s_1, s_2 \rbrace$, $a_{1}, b_{1} \in \lbrace L, D, R, U \rbrace$, $a_{2}, b_{2} \in \lbrace \uparrow, \downarrow \rbrace$. The coefficients $\kappa^{s_1 L \downarrow}_{s_2 R \downarrow}=-\kappa^{s_1 D \uparrow}_{s_2 U \uparrow}=\frac{1}{4} \left( \kappa_{1} + \kappa_{2} \right)$, $\kappa^{s_1 D \uparrow}_{s_2 R \downarrow}=\kappa^{s_1 L \downarrow}_{s_2 U \uparrow}=\frac{i}{4} \left( \kappa_{1} - \kappa_{2} \right)$, $\kappa^{s_2 R \downarrow}_{s_2 R \downarrow}=\kappa^{s_2 U \uparrow}_{s_2 U \uparrow}=\frac{i \kappa_{1}}{4} $, $\kappa^{s_1 D \uparrow}_{s_1 D \uparrow}=\kappa^{s_1 L \downarrow}_{s_1 L \downarrow}=\frac{i \kappa_{2}}{4}$ and the rest are 0, where $\kappa_{1}=e^{ i \left( \phi_{R} + \phi_{L}\right)} $ and $\kappa_{2}= e^{ i \left( \phi_{D} + \phi_{U}\right)}$ (see Supplemental Material~\cite[Appendix D]{Supply}).

Next we show maximal violation of MoE through squared concurrence measure as follows. First, we apply a projector $\Pi_{s_1s_2}$ on $\ket{\Psi^{(2)}}$ as in~\cite{LFC18,Nosrati19}, so that Alice and Bob have exactly one particle each, where 
\frenchspacing
 \medmuskip=-1mu
\thinmuskip=-1mu 
\thickmuskip=-1mu 
\begin{equation}
\Pi_{s_1s_2} := \sum_{ \zeta_{1} \in \lbrace L,D \rbrace, \zeta_{2} \in \lbrace R,U \rbrace, \tau \in \lbrace \downarrow, \uparrow \rbrace}  \ket{s_1 \zeta_{1} \tau, s_2 \zeta_{2} \tau} \bra{s_1 \zeta_{1} \tau, s_2 \zeta_{2} \tau}.
\end{equation}
\frenchspacing
 \medmuskip=2mu
\thinmuskip=2mu
\thickmuskip=2mu 
This results in the normalized density matrix $\rho^{(2)}_{s_1s_2}=\ket{\Psi^{(2)}}_{s_1s_2}\bra{\Psi^{(2)}}$ where $\ket{\Psi^{(2)}}_{s_1s_2}= \frac{\Pi_{s_1s_2} \ket{\Psi^{(2)}} }{\sqrt{\braket{\Psi^{(2)}|\Pi_{s_1s_2}|\Psi^{(2)}}}}$.
Now, we have to trace-out the path DoF of each particle from $\rho^{(2)}_{s_1s_2}$ using Eq.~\eqref{DoFTrRule} resulting in the reduced density matrix $\rho_{s_{1_{a_2}}s_{2_{b_2}}}$. 

Calculations show that the concurrence is $\mathcal{C}_{s_{1_{a_2}} \mid s_{2_{b_2}}}( \rho_{s_{1_{a_2}}s_{2_{b_2}}})=1$, when Alice and Bob both measure the particles in spin DoF (see Supplemental Material~\cite[Appendix D]{Supply}). Similarly, the concurrence $\mathcal{C}_{s_{1_{a_2}} \mid s_{2_{b_1}}}( \rho_{s_{1_{a_2}}s_{2_{b_1}}})$ when Alice measures in spin DoF and Bob measures in path DoF is also 1. Thus we clearly get a maximum violation of Eq.~\eqref{Degree_Monogamy}. 

Interestingly, this violation is irrespective of any particular entanglement measure like squared concurrence. It can be shown that such a violation happens in indistinguishable particles by any monogamous bipartite entanglement measure for qubit systems. This leads to the following result.
\begin{theorem}\label{thm1}
{In qubit systems, indistinguishability is a necessary criterion for maximum violation of monogamy of  entanglement by the same measures that are monogamous for distinguishable particles}.
\end{theorem}
For a formal proof of this theorem, please see Supplemental Material~\cite[Appendix E]{Supply}.

\emph{Discussion.--}
MoE is widely regarded as one of the basic principles of quantum physics~\cite{Terhal04}. Qualitatively, it is always expected to hold, as a maximal violation will have consequences for the no-cloning theorem. So much so, that a quantitative violation is interpreted as the non-monogamistic nature of the entanglement measure and not of the system of particles itself~\cite{Lancien16}. Some of those non-monogamous measures can be elevated to be monogamous through convex roof extension~\cite{Kim09}.

Through Theorem 1, we establish a qualitative violation of MoE which was hitherto unheard of. We show that using quantum indistinguishability, it is possible to maximally violate the MoE for all such entanglement measures which are known to be monogamous for distinguishable systems. To establish this theorem, we first needed to modify the qualitative definition of MoE itself, transiting from the particle-view to the DoF-view and had to introduce the DoF trace-out rule for indistinguishable particles. Thus, this is a non-trivial extension of the well-known MoE.

Further, our framework also takes into account the recently introduced inter-DoF entanglement~\cite{HHNL}.  Quantum physics dictates the measurement results of a particular DoF when it is correlated with another DoF. Taking a partial trace while keeping the rule of quantum physics intact is extremely non-trivial and requires a rigorous mathematical treatment. Our framework, therefore, captures these nuances of quantum physics better than any other existing framework.

Theorem~\ref{thm1} unveils a non-trivial difference between distinguishable and indistinguishable systems. For distinguishable systems, MoE and no-cloning theorem imply one another (see Supplemental Material~\cite[Appendix B]{Supply}).
 The significance of our result is that for indistinguishable systems, the no-cloning theorem remains more fundamental than MoE and the former does not necessarily imply the latter. In fact, no-cloning is derived from the linearity of quantum mechanics~\cite{Wootters82} and hence even indistinguishable particles are also bound to follow it. It appears that the only way to reconcile the co-existence of MoE violation and no-cloning for indistinguishable particles is to consider that such particles do not yield unit fidelity in quantum teleportation~\cite{Ugo15,LFC18,Das20}.

Moreover, indistinguishability is not a sufficient condition for violation of MoE. There can be scenarios where indistinguishable subsystems may be maximally entangled, respecting monogamy. Only specific entanglement structures (such as the circuit discussed in this work) can lead to a maximum violation of MoE. That is why we call indistinguishability a necessary criterion for maximum violation of MoE.

Theorem~\ref{thm1} raises a few fundamental questions on the properties of entanglement for indistinguishable particles. 

(i) There are several applications of MoE for distinguishable particles such as~\cite{Pawlowski10,Tomamichel13,Johnston16,Dur00,Bae06,Chiribella06,Lloyd14}. In particular, for cryptographic applications~\cite{Pawlowski10,Tomamichel13,Johnston16}, MoE provides security in the distinguishable scenario. What happens to such applications in the indistinguishable case?

(ii) Can there be a new application of sharability of maximal entanglement among indistinguishable particles that are not possible for distinguishable ones? 

(iii) Do indistinguishable particles also exhibit maximum violation of monogamy for general quantum correlations~\cite{Streltsov12} such as discord~\cite{Braga12,Prabhu12,Bai13,Liu14}, coherence~\cite{Radhakrishnan16,Basso20}, steering~\cite{Reid13,Milne15,Cheng16}, etc.?

All the above remains a matter of thorough investigation and can be part of interesting future works. 

\emph{Conclusion.--} 
We have proposed a DoF trace-out rule applicable for both distinguishable and indistinguishable particles. Using this tool, we have established the following counter-intuitive result: MoE can be violated maximally for indistinguishable particles by any bipartite entanglement measure that is known to be monogamous for distinguishable particles. MoE, in essence, is a no-go theorem that is a restriction on the shareability of entanglement. Our result lifts this restriction for indistinguishable particles. It also opens up a new area where researchers can investigate whether the  applications of MoE using distinguishable particles are also applicable using indistinguishable ones and their advantages and disadvantages.

\section{Supplemental Material}

\subsection{Appendix A: Revisiting the notation of Lo Franco \textit{et al.}~\cite{LFC16} for indistinguishable particles} \label{AppA}
If the state vector of two indistinguishable particles are labeled by $\phi$ and $\psi$, then the two-particle state is represented by a single entity $\ket{\phi,\psi}$. The  two-particle probability amplitudes is represented by
\begin{equation}\label{inner_P}
\braket{\varphi,\zeta|\phi,\psi} := \braket{\varphi|\phi}\braket{\zeta|\psi} + \eta \braket{\varphi|\psi}\braket{\zeta|\phi},
\end{equation}
where $\varphi,\zeta$ are one-particle states of another global two-particle state vector and $\eta = 1$ for bosons and $\eta=-1$ for fermions. The right hand side of Eq.~\eqref{inner_P} is symmetric if one-particle state position
is swapped with another, i.e., $\ket{\phi,\psi}=\eta \ket{\psi,\phi}$.
From Eq.~\eqref{inner_P}, the probability of finding two particles in the same state $\ket{\varphi}$ is $\braket{\varphi,\varphi|\phi,\psi} = (1+\eta) \braket{\varphi|\phi}\braket{\varphi|\psi}$ which is zero for fermions due to Pauli exclusion principle~\cite{Pauli25} and maximum for bosons. 
As Eq.~\eqref{inner_P} follows symmetry and linearity property, the symmetric inner product of states
with spaces of different dimensionality is defined as
\begin{equation}\label{SinplePunN}
\bra{\psi_{k}}\cdot \ket{\varphi_{1},\varphi_{2}} \equiv \braket{\psi_{k} \mid \varphi_{1},\varphi_{2}} = \braket{\psi_{k}|\varphi_{1}}\ket{\varphi_{2}} + \eta \braket{\psi_{k}|\varphi_{2}}\ket{\varphi_{1}},
\end{equation} 
where $\ket{\tilde{\Phi}}=\ket{\varphi_{1},\varphi_{2}}$ is the un-normalized state of two indistinguishable particles and $\ket{\psi_{k}}$ is a single-particle state. Equation~\eqref{SinplePunN} can be interpreted as a projective measurement where the two-particle un-normalized state $\ket{\tilde{\Phi}}$ is projected into a single particle state $\ket{\psi_{k}}$. Thus, the resulting normalized pure-state of a single particle after the projective measurement can be written as
\begin{equation}
\ket{\phi_{k}}=\frac{\braket{\psi_{k}|\Phi}}{\sqrt{\braket{\Pi^{(1)}_{k}}}_{\Phi}}, 
\end{equation}
where $\ket{\Phi}:=\frac{1}{\sqrt{\mathbb{N}}}\ket{\tilde{\Phi}}$ with $\mathbb{N}=1+\eta \mid \braket{\varphi_{1}|\varphi_{2}} \mid^2$ and $\Pi^{(1)}_{k}=\ket{\psi_{k}} \bra{\psi_{k}}$ is the one-particle projection operator. The one-particle identity operator can be defined as $\mathbb{I}^{(1)}:=\sum_{k}\Pi^{(1)}_{k}$. So, using the linearity property of projection operators, one can write similar to Eq.~\eqref{SinplePunN}: 
\begin{equation}
\ket{\psi_{k}} \bra{\psi_{k}} \cdot  \ket{\varphi_{1},\varphi_{2}} =  \braket{\psi_{k}|\varphi_{1}}\ket{\psi_{k},\varphi_{2}} + \eta \braket{\psi_{k}|\varphi_{2}}\ket{\varphi_{1},\psi_{k}}.
\end{equation} 
Note that
\begin{equation} \label{Identity}
\mathbb{I}^{(1)} \ket{\Phi} = 2 \ket{\Phi},
\end{equation}
where the probability of resulting the state $\ket{\psi_{k}}$ is $p_{k}=\braket{\Pi^{(1)}_{k}}_{\Phi}/2$. The partial trace in this method is can be written as 
\begin{equation}\label{tracedef}
\rho^{(1)} = \frac{1}{2} \text{Tr}^{(1)} \ket{\Phi} \bra{\Phi} =\frac{1}{2} \sum_{k} \braket{\psi_{k}|\Phi} \braket{\Phi|\psi_{k}} = \sum_{k} p_{k} \ket{\phi_{k}} \bra{\phi_{k}},
\end{equation}
where the factor $1/2$ comes from Eq.~\eqref{Identity}. 
 
Another useful concept that of \textit{localized partial trace}~\cite{LFC16}, which means that local measurements are being performed on a region of space $M$ where  the particle has a non-zero probability of being found. So, performing the localized partial trace on a region $M$, we get
\begin{equation}
\rho^{(1)}_{M}=\frac{1}{\mathbb{N}_{M}}\text{Tr}^{(1)}_{M} \ket{\Phi} \bra{\Phi},
\end{equation}
where $\mathbb{N}_{M}$  is a normalization constant such that $\text{Tr}^{(1)}\rho^{(1)}_{M}=1$. The entanglement entropy can be calculated as
\begin{equation} \label{IndVNE}
E_{M}(\ket{\Phi}) := S(\rho^{(1)}_{M}) = -\sum_{i} \lambda_{i}\text{ln}\lambda_{i},
\end{equation} 
where $S(\rho)=-\text{Tr}(\rho \text{ln} \rho)$ is the von Neumann entropy and $\lambda_{i}$ are the eigenvalues of $\rho^{(1)}_{M}$. We will call the state as entangled state if we get a non-zero value of Eq.~\eqref{IndVNE}.

\subsection{Appendix B: Equivalence of the monogamy of entanglement and the no-cloning theorem for distinguishable particles} \label{AppB}
\begin{figure*}[htbp]
\centering
\includegraphics[width=0.75\textwidth]{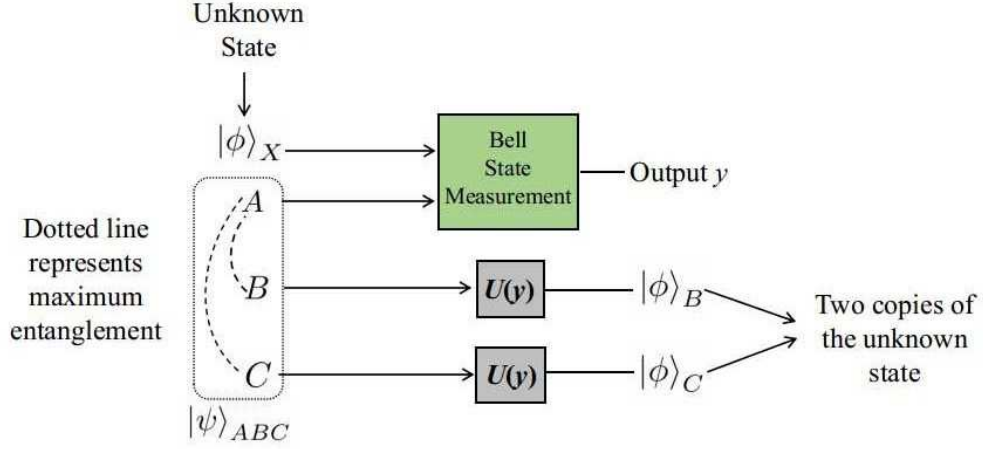} 
\caption{Circuit to get violation of the no-cloning theorem from the maximum violation of MoE.}
\label{fig:BSM}
\end{figure*}
To show that no-cloning implies monogamy of entanglement (MoE), let us prove its contrapositive. When MoE is violated maximally, one can achieve quantum cloning~\cite{QC05,QC14} of any unknown quantum state using standard teleportation protocol~\cite{QT93,QTnat15} as follows. Assume a particle $A$ is maximally entangled with the particles $B$ and $C$ and their joint state is denoted by $\ket{\psi}_{ABC}$ and the particle $X$ is in unknown quantum state $\ket{\phi}_{C}$. To achieve cloning of the state $\ket{\phi}$, one has to perform Bell state measurements (BSM)~\cite{BSM99} jointly on the particles $A$ and $X$. Based on the measurement result denoted by $y$, suitable unitary operations $U_{y}$ have to be performed on the particles $B$ and $C$ so that the state $\ket{\phi}$ appears on each of them, where $U_{y} \in \left\lbrace \mathcal{I}, \sigma_{x}, \sigma_{y}, \sigma_{z}\right\rbrace $, $\mathcal{I}$ being the identity operation and $\sigma_{i}$'s $\left( i = x, y, z \right)$ the Pauli matrices. Thus we can have two copies of the unknown state $\ket{\phi}$ as $\ket{\phi}_{B}$ and $\ket{\phi}_{C}$. 

Next, to show that MoE implies no-cloning, again we prove its contrapositive. Let two particles $A$ and $B$ share a maximally entangled state $\ket{\psi}_{AB}$. If possible, suppose one of them, say, $B$ is cloned and we get a copy $B_{1}$ of $B$, then in the tripartite state $\ket{\psi}_{ABB_{1}}$, $A$ is maximally entangled with both $B$ and $B_{1}$ simultaneously, thus violating the MoE maximally.

\subsection{Appendix C: Degree of freedom (DoF) trace-out rule for two indistinguishable particles} \label{AppC}
Lo Franco \textit{et al.}~\cite{LFC16} have defined the partial trace-out rule for indistinguishable particles where each particle has a spatial label and a single DoF. By a non-trivial extension of their concept, We define DoF trace-out rule when each indistinguishable particle has $n$ DoFs and a spatial label. 
Lets us consider two indistinguishable particles each having $n$ DoFs, having spatial labels  $\alpha$ and $\beta$ whose $i$-th and $j$-th DoF are represented by $a_{i}$ and $b_{j}$ respectively, where $i, j \in \mathcal{N} =\lbrace 1, 2, \cdots , n \rbrace$. The general state of two particles can be represented by 
\begin{equation} \label{State}
\begin{aligned}
\ket{\Psi^{(n)}}= \sum_{\alpha, \beta, a_{1}, a_{2}, \cdots , a_{n},  b_{1}, b_{2}, \cdots, b_{n}}\kappa^{\alpha a_{1} a_{2} \cdots a_{n}}_{\beta  b_{1} b_{2} \cdots b_{n}} \ket{\alpha a_{1} a_{2} \cdots a_{n}, \beta  b_{1} b_{2} \cdots b_{n} }.
\end{aligned}
\end{equation}
Here, $  \alpha , \beta$ ranges over $ \mathcal{S} = \left\lbrace s_{1}, s_{2}, \cdots , s_{p} \right \rbrace $ which refers to distinct spatial modes with $p \geq 2$. Each of  $a_{i}, b_{i}$ ranges over $ \mathcal{D}_{i} = \left\lbrace D_{i_{1}}, D_{i_{2}}, \cdots , D_{i_{q_{i}}} \right\rbrace $ which refers to the eigenvalues of the $i$-th  DoF, where $i \in \mathcal{N} $. Note that $q_{i} \geq 2$ since each DoF must have atleast two distinct values. The value of $q_{i}$ may be different for different $i$.

 The general density matrix of two indistinguishable particles can be described as
 \medmuskip=0mu
\thinmuskip=0mu
\thickmuskip=0mu
\begin{equation} \label{DM}
\begin{aligned}
\rho^{(n)}= \sum_{\alpha, \beta,  a_{1}, a_{2}, \cdots, a_{n},  b_{1}, b_{2}, \cdots, b_{n} } \hspace{0.3cm} \sum_{\gamma, \delta, c_{1}, c_{2}, \cdots, c_{n},   d_{1}, d_{2}, \cdots, d_{n}} \kappa^{\alpha a_{1} a_{2} \cdots a_{n} }_{\beta  b_{1} b_{2} \cdots b_{n}} \kappa^{\gamma c_{1} c_{2} \cdots c_{n} *}_{\delta  d_{1} d_{2} \cdots d_{n}}  \ket{\alpha a_{1} a_{2} \cdots a_{n}, \beta  b_{1} b_{2} \cdots b_{n}} \bra{\gamma c_{1} c_{2} \cdots c_{n}, \delta  d_{1} d_{2} \cdots d_{n}},
\end{aligned}
\end{equation}
 \medmuskip=2mu
\thinmuskip=2mu
\thickmuskip=2mu
where $ \alpha, \beta ,\gamma , \delta$ span $ \mathcal{S} $ and $a_{i}, b_{i}, c_{i}, d_{i}$ span $ \mathcal{D}_{i}$ with $i\in  \mathcal{N}$. If we want to perform partial trace in only one region, say $s_{x} \in \mathcal{S}$, then the non-normalized density matrix can be written as
\begin{equation} \label{Trace2p_Simple}
\begin{aligned}
\tilde{\rho}^{(1)}_{M} = \text{Tr}_{M} \left( \rho^{(n)} \right) =\sum_{m_{1},m_{2}, \cdots, m_{n}} \braket{s_{x} m_{1} m_{2} \cdots m_{n} | \rho^{(n)} | s_{x} m_{1} m_{2} \cdots m_{n}},
\end{aligned}
\end{equation}
where $m_{i}$ span $ \mathcal{D}_{i}$, where $i\in  \mathcal{N}$. Equations~\eqref{State} to~\eqref{Trace2p_Simple} can trivially be generalized for $n$ particles. 

We define the probability of the joint measurement on $\ket{\Psi^{(n)}}$, where one particle is measured in the $h$-th eigenvalue of the $i$-th DoF denoted by $D_{i_{h}}$  in localized region $s_{x}$ and another particle is measured in the $k$-th eigen value of $j$-th DoF denoted by $D_{j_{k}}$ in localized region $s_{y}$ where $x , y \in \lbrace 1, 2, \cdots , p \rbrace$, $i, j \in \mathcal{N}$, $h \in \lbrace 1, 2, \cdots, q_{i} \rbrace$, and $k \in \lbrace 1, 2, \cdots, q_{j} \rbrace$, as below: 
 \begin{equation} \label{TracenDoF}
\begin{aligned}
& \sum_{\alpha, \beta, a_{1}, a_{2}, \cdots , a_{n},  b_{1}, b_{2}, \cdots, b_{n}} \braket{s_{x} a_{1} a_{2} \cdots a_{i-1} D_{i_{h}}  a_{i+1} \cdots,  a_{n}, s_{y} b_{1} b_{2} \cdots b_{j-1} D_{j_{k}} b_{j-1} \cdots  b_{n}  |\Psi^{(n)}}.
\end{aligned}
\end{equation}
The joint measurement rule defined in Eq.~\eqref{TracenDoF} is the generalized version of the joint measurement rule defined in~\eqref{inner_P}.

Now we define the DoF trace-out rule. First, we consider distinguishable particles, and then we will extend it for indistinguishable particles.

If the particles $A$ and $B$ were distinguishable such that the particle $A$ and $B$ are in the spatial region $\alpha$ and $\beta$ respectively, then Eq.~\eqref{State} could be represented as
\begin{equation} \label{Statedis}
\begin{aligned}
\ket{\Psi^{(n)}}_{AB}= \sum_{a_{1}, a_{2}, \cdots , a_{n},  b_{1}, b_{2}, \cdots, b_{n}}\kappa^{ a_{1} a_{2} \cdots a_{n}}_{  b_{1} b_{2} \cdots b_{n}} \ket{ a_{1} a_{2} \cdots a_{n}} \otimes \ket{  b_{1} b_{2} \cdots b_{n} },
\end{aligned}
\end{equation}
and the density matrix of Eq.~\eqref{DM} would take the form
\begin{equation} \label{DMdis}
\begin{aligned}
\rho^{(n)}_{AB}= \sum_{a_{1}, a_{2}, \cdots, a_{n},  b_{1}, b_{2}, \cdots, b_{n} } \hspace{0.1cm} \sum_{c_{1}, c_{2}, \cdots, c_{n},   d_{1}, d_{2}, \cdots, d_{n}} \kappa^{ a_{1} a_{2} \cdots a_{n} }_{  b_{1} b_{2} \cdots b_{n}} \kappa^{ c_{1} c_{2} \cdots c_{n} *}_{  d_{1} d_{2} \cdots d_{n}}  \ket{ a_{1} a_{2} \cdots a_{n}} \ket{  b_{1} b_{2} \cdots b_{n}} \otimes \bra{ c_{1} c_{2} \cdots c_{n}}\bra{  d_{1} d_{2} \cdots d_{n}}.
\end{aligned}
\end{equation}

 If we want to trace-out the $i$-th DoF of particle $A$, then from Eq.~\eqref{DMdis}, the reduced density matrix can be written as
\begin{equation} \label{DisDoFTraceout}
\begin{aligned}
\rho_{a_{\bar{i}}} \equiv & \text{Tr}_{a_{i}} \left( \rho^{(n)}_{AB} \right) := \sum_{ a_{i}, a_{\bar{i}},  b_{1}, b_{2}, \cdots, b_{n} } \hspace{0.2cm} \sum_{ c_{i} c_{\bar{i}},d_{2}, \cdots, d_{n}} \kappa^{ a_{\bar{i}}}_{ b_{1} b_{2} \cdots b_{n} } \kappa^{ c_{\bar{i}} *}_{ d_{1} d_{2} \cdots d_{n}}  \ket{ a_{\bar{i}}} \ket{   b_{1} b_{2} \cdots b_{n} }  \bra{c_{\bar{i}}} \bra{ d_{1} d_{2} \cdots d_{n}}
 \left\lbrace \braket{a_{i}|c_{i}} \right\rbrace ,
\end{aligned}
\end{equation}
where $a_{\bar{i}}=a_{1}a_{2} \cdots a_{i-1}a_{i+1} \cdots a_{n}$ and similar meaning for $c_{\bar{i}}$. One can show that when the DoF trace-out rule in Eq.~\eqref{DisDoFTraceout} is applied to the same particle for $n$ times, it becomes equivalent to our familiar particle trace-out rule~\cite[Eq. 2.178]{Nielsenbook}.

Next we  define DoF trace-out rule for indistinguishable particles from the general density matrix of two particles as defined in Eq.~\eqref{DM}. Suppose we want to trace-out the $i$-th DoF of location $s_x \in \mathcal{S}$. Then the DoF reduced density matrix is
 \medmuskip=0mu
\thinmuskip=0mu
\thickmuskip=0mu
\begin{equation} \label{DOFTrRule}
\begin{aligned}
\rho_{s_{x_{\bar{i}}}} \equiv  \text{Tr}_{s_{x_{i}}} \left(  \rho^{(n)} \right) \equiv & \sum_{m_{i}} \braket{ s_x m_i \mid \rho^{(n)} \mid s_x m_i } \\
:=& \sum_{m_{i}} \bigg \lbrace \sum_{\substack{ \alpha, \beta, a_{i}, a_{\bar{i}},  b_{1}, b_{2}, \cdots, b_{n}, \\ \gamma, \delta, c_{i}, c_{\bar{i}}, d_{1}, d_{2}, \cdots, d_{n}} } \kappa^{\alpha a_{i} a_{\bar{i}}}_{\beta b_{1} b_{2} \cdots b_{n}} \kappa^{\gamma c_{i} c_{\bar{i}} *}_{\delta d_{1} d_{2} \cdots d_{n}}  \braket{s_x m_i \mid \alpha a_i} \braket{\gamma c_i  \mid s_x m_i}  \ket{\alpha a_{\bar{i}}, \beta  b_{1} b_{2} \cdots b_{n}}\bra{\gamma c_{\bar{i}}, \delta  d_{1} d_{2} \cdots d_{n}}  \\
& +  \eta \sum_{\substack{ \alpha, \beta,  a_{1}, a_{2}, \cdots, a_{n}, b_{i}, b_{\bar{i}}, \\ \gamma, \delta, c_{i}, c_{\bar{i}}, d_{1}, d_{2}, \cdots, d_{n}} } \kappa^{\alpha a_{1} a_{2} \cdots a_{n}}_{\beta b_{i} b_{\bar{i}}} \kappa^{\gamma c_{i} c_{\bar{i}} *}_{\delta d_{1} d_{2} \cdots d_{n}} \braket{s_x m_i \mid \beta b_i }  \braket{\gamma c_i  \mid s_x m_i}  \ket{\alpha a_{1} a_{2} \cdots a_{n}, \beta  b_{\bar{i}}} \bra{\gamma c_{\bar{i}}, \delta  d_{1} d_{2} \cdots d_{n}} 
\\
&+ \eta  \sum_{\substack{ \alpha, \beta, a_{i}, a_{\bar{i}},  b_{1}, b_{2}, \cdots, b_{n}, \\ \gamma, \delta, c_{1}, c_{2}, \cdots, c_{n} , d_{i}, d_{\bar{i}}} } \kappa^{\alpha a_{i} a_{\bar{i}}}_{\beta b_{1} b_{2} \cdots b_{n}} \kappa^{\gamma c_{1} c_{2} \cdots c_{n} *}_{\delta d_{i} d_{\bar{i}}}  \braket{s_x m_i \mid \alpha a_i}  \braket{\delta d_i  \mid s_x m_i}  \ket{\alpha a_{\bar{i}}, \beta  b_{1} b_{2} \cdots b_{n}}\bra{\gamma c_{1} c_{2} \cdots c_{n}, \delta  d_{\bar{i}}}   \\
 & + \sum_{\substack{ \alpha, \beta,  a_{1}, a_{2}, \cdots, a_{n}, b_{i}, b_{\bar{i}}, \\ \gamma, \delta, c_{1}, c_{2}, \cdots, c_{n} , d_{i}, d_{\bar{i}}} } \kappa^{\alpha a_{1} a_{2} \cdots a_{n}}_{\beta b_{i} b_{\bar{i}}} \kappa^{\gamma c_{1} c_{2} \cdots c_{n} *}_{\delta d_{i} d_{\bar{i}}}  \braket{s_x m_i \mid \beta b_i } \braket{\delta d_i  \mid s_x m_i}   \ket{\alpha a_{1} a_{2} \cdots a_{n}, \beta  b_{\bar{i}}} \bra{\gamma c_{1} c_{2} \cdots c_{n}, \delta  d_{\bar{i}}} \bigg \rbrace ,
\end{aligned}
\end{equation}
 \medmuskip=2mu
\thinmuskip=2mu
\thickmuskip=2mu
where $a_{\bar{i}} := a_{1}a_{2} \cdots a_{i-1}a_{i+1} \cdots a_{n}$ and similar meaning for  $b_{\bar{i}}$, $c_{\bar{i}}$ and $d_{\bar{i}}$. 

It may be noted that for $n=2$, the DoF trace-out rule defined in Eq.~\eqref{DOFTrRule} reduces to Eq. (6) in the main text. For $n=1$,  this becomes equivalent to the  particle trace-out rule as defined Eq.~\eqref{tracedef}. On the other hand, for $n > 1$, if we apply DoF trace-out rule of Eq.~\eqref{DOFTrRule} $n$ times, the effect will not be the same as the particle trace-out in Eq.~\eqref{tracedef}. The reason behind this is as follows. For indistinguishable particles, the particle trace-out operation vanishes all the DoFs together for one particle; whereas each DoF trace-out operation leaves an expression with many terms each of which vanishes the corresponding DoF from one particle at a time and retains the same DoF in the remaining particles.

\subsection{Appendix D: Description of the circuit of Li \textit{et al.}~\cite{HHNL}}\label{AppD}
The circuit of Yurke \textit{et al.}~\cite{Y&S92PRA,Y&S92PRL} to generate  quantum entanglement between the same DoFs of two indistinguishable particles (bosons and fermions) is extended by Li \textit{et al.}~\cite{HHNL} to generate inter-DoF entanglement between two indistinguishable bosons. Details of their generation scheme are as follows.

For bosons, the second quantization formulation deals with bosonic operators $b_{i,\textbf{p}}$ with $\ket{i,\textbf{p}}=b^{\dagger}_{i,\textbf{p}}\ket{0}$, where $\ket{0}$ is the vacuum and $\ket{i,\textbf{p}}$ describes a particle with spin $\ket{i}$ and momentum $\textbf{p}$. These operators satisfy the canonical commutation relations:

\begin{equation}
\left[  b_{i,\textbf{p}_{i}}, b_{j,\textbf{p}_{j}} \right]  = 0, \hspace{0.2cm} \left[  b_{i,\textbf{p}_{i}}, b^{\dagger}_{j,\textbf{p}_{j}}\right]  = \delta(\textbf{p}_{i}-\textbf{p}_{j})\delta_{ij}.
\end{equation} 

\begin{figure*}[t!]
\centering
\includegraphics[width=\textwidth]{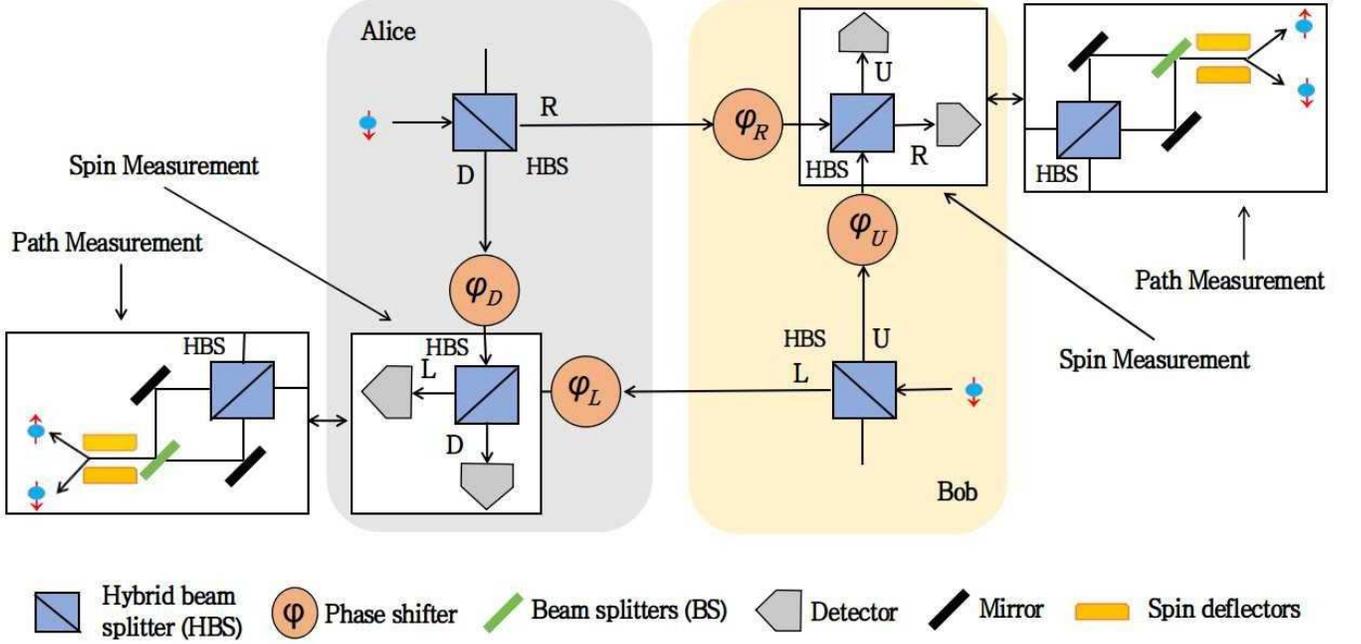} 
\caption{Circuit to generate hyper-hybrid entangled state as proposed by Li \textit{et al.}~\cite{HHNL}. Here the bi-directional arrow represents the measurement is done either in spin DoF or in Path DoF.}
\label{fig:Lietal}
\end{figure*} 

Analysis of the circuit of Li \textit{et al.}~\cite{HHNL} for bosons involves an array of hybrid beam splitters (HBS)~\cite[Fig. 3]{HHNL}, phase shifts, four orthogonal external modes $L$, $D$, $R$ and $U$ and two orthogonal internal modes $\uparrow$ and $\downarrow$ as shown in Fig.~\ref{fig:Lietal}. Here, particles exiting through the modes $L$ and $D$ are received by Alice (A) who can control the phases $\varphi_{L}$ and $\varphi_{D}$, whereas particles exiting through the modes $R$ and $U$  are received by Bob (B) who can control the phases $\varphi_{R}$ and $\varphi_{U}$. 

In this circuit, two particles, each with spin $\ket{\downarrow}$, enter the set up in the mode $R$ and $L$ for Alice and Bob respectively. The initial state of the two particles is $\ket{\Psi_{0}}=b^{\dagger}_{\downarrow,R} b^{\dagger}_{\downarrow,L}\ket{0}$. Now, the particles are sent to HBS such that one output port of HBS is sent to other party ($R$ or $L$) and the other port remains locally accessible ($D$ or $U$). Next, each party applies state-dependent (or spin-dependent) phase shifts. Lastly, the output of local mode and that received from the other party is mixed with HBS and then the measurement is performed in either  external or internal modes. The final state can be written as
\begin{equation} \label{Final_state_boson}
\begin{aligned}
&\ket{\Psi}=\frac{1}{4} \left[ e^{i\varphi_{R}}\left( b^{\dagger}_{\downarrow,R}+ib^{\dagger}_{\uparrow,U}\right) +ie^{i\varphi_{D}} \left( b^{\dagger}_{\uparrow,D} + i b^{\dagger}_{\downarrow,L}\right)  \right] 
 \otimes \left[ e^{i\varphi_{L}} \left( b^{\dagger}_{\downarrow,L}+ib^{\dagger}_{\uparrow,D}\right)  + i e^{i\varphi_{U}}\left( b^{\dagger}_{\uparrow,U}+ib^{\dagger}_{\downarrow,R}\right) \right] \ket{0}. 
\end{aligned}
\end{equation}

Now we represent the circuit of Li \textit{et al.} using our proposed extended version of the notation of Lo Franco \textit{et al.} as described in Appendix C. The final state in Eq.~\eqref{Final_state_boson} can be represented by Eq.~\eqref{State} as 
\begin{equation} \label{Li2Lo}
\ket{\Psi^{(2)}}= \sum_{\alpha, \beta \in \lbrace s_1, s_2 \rbrace, a_{1}, b_{1} \in \lbrace L, D, R, U \rbrace, a_{2}, b_{2} \lbrace \uparrow, \downarrow \rbrace}  \kappa^{\alpha a_{1} a_{2}}_{\beta  b_{1} b_{2}} \ket{\alpha a_{1} a_{2}, \beta  b_{1} b_{2}},
\end{equation}
where the coefficients are
\begin{equation*}
\kappa^{s_1 L \downarrow}_{s_2 R \downarrow}=-\kappa^{s_1 D \uparrow}_{s_2 U \uparrow}=\frac{1}{4} \left( \kappa_{1} + \kappa_{2} \right), \hspace{0.2cm} \kappa^{s_1 D \uparrow}_{s_2 R \downarrow}=\kappa^{s_1 L \downarrow}_{s_2 U \uparrow}=\frac{i}{4} \left( \kappa_{1} - \kappa_{2} \right), \hspace{0.2cm} \kappa^{s_2 R \downarrow}_{s_2 R \downarrow}=\kappa^{s_2 U \uparrow}_{s_2 U \uparrow}=\frac{i \kappa_{1}}{4}, \hspace{0.2cm} \kappa^{s_1 D \uparrow}_{s_1 D \uparrow}=\kappa^{s_1 L \downarrow}_{s_1 L \downarrow}=\frac{i \kappa_{2}}{4}, 
\end{equation*}
  and the rest are 0, where $\kappa_{1}=e^{ i \left( \phi_{R} + \phi_{L}\right)} $ and $\kappa_{2}= e^{ i \left( \phi_{D} + \phi_{U}\right)}$. Here, we denote the specialized location where Alice and Bob have performed the measurement as $s_1$ and $s_2$ respectively.

\subsection{Appendix E: Proof of Theorem 1}\label{AppE}
In this Section, we will prove our Theorem 1 of the main text. We will give details calculations that maximum violation of MoE happens using squared concurrence measure and log-negativity as entanglement measure. The similar calculation will follow for other monogamous entanglement measures also.

First, we calculate concurrence of the state $\ket{\Psi^{(2)}}$ as described in Eq.~\eqref{Li2Lo} of Appendix D.	

Projecting $\rho^{(2)}=\ket{\Psi^{(2)}}\bra{\Psi^{(2)}}$ onto the (operational) subspace spanned by the computational basis 
\begin{equation}
\begin{aligned}
\Omega_{s_1s_2}= \lbrace & \ket{ s_1 L \downarrow, s_2 R \downarrow},  \ket{ s_1 L \downarrow, s_2 U \downarrow},  \ket{s_1 L \downarrow, s_2 R \uparrow},  \ket{s_1 L \downarrow, s_2 U \uparrow} 
, \ket{s_1 L \uparrow, s_2 R \downarrow},  \ket{s_1 L \uparrow, s_2 U \downarrow},  \ket{s_1 L \uparrow, s_2 R \uparrow},   \\
& \ket{s_1 L \uparrow, s_2 U \uparrow},\ket{s_1 D \downarrow, s_2 R \downarrow},  \ket{s_1 D \downarrow, s_2 U \downarrow},  \ket{s_1 D \downarrow, s_2 R \uparrow},  \ket{s_1 D \downarrow, s_2 U \uparrow} 
, \ket{s_1 D \uparrow, s_2 R \downarrow},  \ket{s_1 D \uparrow, s_2 U \downarrow},\\
& \ket{s_1 D \uparrow, s_2 R \uparrow},  \ket{s_1 D \uparrow, s_2 U \uparrow}
 \rbrace, 
\end{aligned}
\end{equation}
by the projector
\begin{equation}
\begin{aligned}
\Pi_{s_1s_2}= & \sum_{\sigma,\tau = \left\lbrace  \uparrow, \downarrow \right\rbrace , \varsigma = \left\lbrace L, D\right\rbrace,  \upsilon = \left\lbrace  R, U \right\rbrace  } \ket{s_1 \varsigma \sigma, s_2 \upsilon \tau}\bra{s_1 \varsigma \sigma, s_2 \upsilon \tau}, 
\end{aligned}
\end{equation}
one gets the distributed resource state where each localized region $s_1$ and $s_2$ have exactly one particle as
\begin{equation}
\begin{aligned}
\ket{\Psi^{(2)}}_{s_1s_2}= \frac{\Pi_{s_1s_2} \ket{\Psi^{(2)}} }{\sqrt{\braket{\Psi^{(2)}|\Pi_{s_1s_2}|\Psi^{(2)}}}} = \sum_{ a_{1} \in \lbrace L, D \rbrace ,  b_{1} \in \lbrace R, U \rbrace , a_{2}, b_{2} \in \lbrace \uparrow, \downarrow \rbrace} \kappa^{s_1 a_{1} a_{2}}_{s_2  b_{1} b_{2}} \ket{s_1 a_{1} a_{2}, s_2  b_{1} b_{2}},
\end{aligned}
\end{equation}
where the non-zero coefficients are
\begin{equation}	
\kappa^{s_1 L \downarrow}_{s_2 R \downarrow}=- \kappa^{s_1 D \uparrow}_{s_2 U \uparrow}=\frac{1}{2 \sqrt{2}} \left(\kappa_{1} + \kappa_{2} \right), \hspace{0.2cm} \kappa^{s_1 D \uparrow}_{s_2 R \downarrow}=\kappa^{s_1 L \downarrow}_{s_2 U \uparrow}=\frac{i}{2 \sqrt{2}} \left(\kappa_{1} - \kappa_{2} \right).
\end{equation}
 The density matrix $\rho^{(2)}_{s_1s_2}=\ket{\Psi^{(2)}}_{s_1s_2}\bra{\Psi^{(2)}}$ can also be calculated as
\begin{equation} \label{rhoAB}
\rho^{(2)}_{s_1s_2}= \dfrac{\Pi_{s_1s_2} \rho^{(2)} \Pi_{s_1s_2}}{\text{Tr}\left( \Pi_{s_1s_2} \rho^{(2)}\right)},
\end{equation}
where $\rho^{(2)}=\ket{\Psi^{(2)}}\bra{\Psi^{(2)}}$.

Now from Eq.~\eqref{rhoAB}, if we trace-out the path DoFs of location $s_1$ and $s_2$ using Eq.~\eqref{DOFTrRule} (the order does not matter), we get the reduced density matrix as
\begin{equation} \label{Rho_s1a2_s2b2}
\begin{aligned}
\rho_{s_{1_{a_{2}}} s_{2_{b_{2}}}}=\text{Tr}_{s_{1_{a_{1}}} s_{2_{b_{1}}}} \left( \rho^{(2)}_{s_1s_2} \right) =\sum_{ a_{2},  b_{2}, c_{2}, d_{2}  \in \lbrace \uparrow, \downarrow \rbrace }\kappa^{s_1 a_{2} }_{s_2 b_{2}} \kappa^{s_1 c_{2} *}_{s_2 d_{2}} \ket{s_1 a_{2} ,s_2  b_{2}}\bra{s_1 c_{2} ,s_2  d_{2}}, 
\end{aligned}
\end{equation}
where 
\begin{equation*}
\begin{aligned}
\kappa^{s_1 \downarrow}_{s_2 \downarrow} =- \kappa^{s_1 \uparrow}_{s_2 \uparrow}=\frac{1}{2 \sqrt{2}} \left(\kappa_{1} + \kappa_{2} \right), \hspace{0.2cm}  \kappa^{s_1 \uparrow}_{ s_2 \downarrow} =\kappa^{s_1 \downarrow}_{s_2 \uparrow}  = \frac{i}{2 \sqrt{2}} \left(\kappa_{1} - \kappa_{2} \right),
\end{aligned}
\end{equation*}
and rest are zero where complex conjugates are calculated accordingly.

To calculate the maximum violation using squared concurrence, first we calculate the following:
\begin{equation}
 \widetilde{\rho}_{s_{1_{a_{2}}}  s_{2_{b_{2}}}}=\sigma^{s_1}_{y} \otimes \sigma^{s_2}_{y} \rho^{*}_{s_{1_{a_{2}}}  s_{2_{b_{2}}}} \sigma^{s_1}_{y} \otimes \sigma^{s_2}_{y},
\end{equation}
where $\sigma^{X}_{y}=\ket{X}\bra{X} \otimes \sigma_{y}$, $X  \in \lbrace s_1, s_2 \rbrace $,  $\sigma_{y}$  is Pauli matrix and the asterisk denotes complex conjugation. 
So, the expression becomes
\begin{equation}
\begin{aligned}
\widetilde{\rho}_{s_{1_{a_{2}}} s_{2_{b_{2}}}}=\sum_{ a_{2},  b_{2}, c_{2}, d_{2}  \in \lbrace \uparrow, \downarrow \rbrace }\tilde{\kappa}^{s_1 a_{2} }_{s_2 b_{2}} \tilde{\kappa}^{s_1 c_{2} *}_{s_2 d_{2}} \ket{s_1 a_{2} ,s_2  b_{2}}\bra{s_1 c_{2} ,s_2  d_{2}}, 
\end{aligned}
\end{equation}
where 
\begin{equation*}
\begin{aligned}
& \tilde{\kappa}^{s_1 \downarrow}_{s_2 \downarrow} \tilde{\kappa}^{s_1 \downarrow * }_{s_2 \downarrow} = - \tilde{\kappa}^{s_1 \downarrow}_{s_2 \downarrow} \tilde{\kappa}^{s_1 \uparrow * }_{s_2 \uparrow}=- \tilde{\kappa}^{s_1 \uparrow}_{s_2 \uparrow} \tilde{\kappa}^{s_1 \downarrow * }_{s_2 \downarrow}=\tilde{\kappa}^{s_1 \uparrow}_{s_2 \uparrow} \tilde{\kappa}^{s_1 \uparrow * }_{s_2 \uparrow}  = \frac{1}{2} \cos^{2} \phi, \\
 & \tilde{\kappa}^{s_1 \downarrow}_{s_2 \downarrow} \tilde{\kappa}^{s_1 \uparrow *}_{s_2 \downarrow} = \tilde{\kappa}^{s_1 \downarrow}_{s_2 \downarrow} \tilde{\kappa}^{s_1 \downarrow *}_{s_2 \uparrow} =  - \tilde{\kappa}^{s_1 \uparrow}_{s_2 \uparrow} \tilde{\kappa}^{s_1 \uparrow *}_{s_2 \downarrow} =- \tilde{\kappa}^{s_1 \uparrow}_{s_2 \uparrow} \tilde{\kappa}^{s_1 \downarrow *}_{s_2 \uparrow} = \tilde{\kappa}^{s_1 \uparrow}_{s_2 \downarrow} \tilde{\kappa}^{s_1 \downarrow * }_{s_2 \downarrow} = \tilde{\kappa}^{s_1 \downarrow}_{s_2 \uparrow} \tilde{\kappa}^{s_1 \downarrow * }_{s_2 \downarrow} =\tilde{\kappa}^{s_1 \uparrow}_{s_2 \downarrow} \tilde{\kappa}^{s_1 \uparrow * }_{s_2 \uparrow}  = \tilde{\kappa}^{s_1 \downarrow}_{s_2 \uparrow} \tilde{\kappa}^{s_1 \uparrow * }_{s_2 \uparrow} = \frac{1}{2} \cos \phi \sin \phi, \\
 & \tilde{\kappa}^{ s_1 \uparrow}_{s_2 \downarrow} \tilde{\kappa}^{s_1 \uparrow * }_{s_2 \downarrow}=\tilde{\kappa}^{s_1 \uparrow}_{s_2 \downarrow} \tilde{\kappa}^{s_1 \downarrow * }_{ s_2 \uparrow} = \tilde{\kappa}^{s_1 \downarrow}_{s_2 \uparrow} \tilde{\kappa}^{s_1 \uparrow * }_{s_2 \downarrow} = \tilde{\kappa}^{s_1 \downarrow}_{s_2 \uparrow} \tilde{\kappa}^{s_1 \downarrow * }_{s_2 \uparrow} = \frac{1}{2} \sin^{2} \phi ,
\end{aligned}
\end{equation*}
with $\phi=\frac{1}{2} \lbrace \phi_{D} + \phi_{U} - \phi_{R} -\phi_{L}\rbrace$.
Now we  calculate concurrence as

\begin{equation}
\begin{aligned}
\mathcal{C}_{s_{1_{a_{2}}} \mid s_{2_{b_{2}}}}\left( \rho_{s_{1_{a_{2}}} s_{2_{b_{2}}}}\right)= \text{max}\left\lbrace 0, \sqrt{\lambda_{4}} - \sqrt{\lambda_{3}} - \sqrt{\lambda_{2}} - \sqrt{\lambda_{1}} \right\rbrace , 
\end{aligned}
\end{equation}
where $\lambda_{i}$ are the eigenvalues, in decreasing order, of the non-Hermitian matrix 
\begin{equation}
\begin{aligned}
\mathbb{R}=\rho_{s_{1_{a_{2}}}  s_{2_{b_{2}}}}\widetilde{\rho}_{s_{1_{a_{2}}}  s_{2_{b_{2}}}}=\sum_{ a_{2},  b_{2}, c_{2}, d_{2}  \in \lbrace \uparrow, \downarrow \rbrace }\bar{\kappa}^{s_1 a_{2} }_{s_2 b_{2}} \bar{\kappa}^{s_1 c_{2} *}_{s_2 d_{2}} \ket{s_1 a_{2} ,s_2 b_{2}}\bra{s_1 c_{2} , s_2 d_{2}}, 
\end{aligned}
\end{equation}
where 
\begin{equation*}
\begin{aligned}
& \bar{\kappa}^{s_1 \downarrow}_{s_2 \downarrow} \bar{\kappa}^{s_1 \downarrow * }_{ s_2 \downarrow} = - \bar{\kappa}^{s_1 \downarrow}_{s_2 \downarrow} \bar{\kappa}^{s_1 \uparrow * }_{s_2 \uparrow}=- \bar{\kappa}^{s_1 \uparrow}_{s_2 \uparrow} \bar{\kappa}^{s_1 \downarrow * }_{s_2 \downarrow}=\bar{\kappa}^{s_1 \uparrow}_{s_2 \uparrow} \bar{\kappa}^{s_1 \uparrow * }_{s_2 \uparrow}  = \frac{1}{4} \cos^{2} \phi, \\
 & \bar{\kappa}^{s_1 \downarrow}_{s_2 \downarrow} \bar{\kappa}^{s_1 \uparrow *}_{s_2 \downarrow} = \bar{\kappa}^{s_1 \downarrow}_{s_2 \downarrow} \bar{\kappa}^{s_1 \downarrow *}_{s_2 \uparrow} =  - \bar{\kappa}^{s_1 \uparrow}_{s_2 \uparrow} \bar{\kappa}^{s_1 \uparrow *}_{s_2 \downarrow} =- \bar{\kappa}^{s_1 \uparrow}_{s_2 \uparrow} \bar{\kappa}^{s_1 \downarrow *}_{s_2 \uparrow} = \bar{\kappa}^{s_1 \uparrow}_{s_2 \downarrow} \bar{\kappa
}^{s_1 \downarrow * }_{s_2 \downarrow} = \bar{\kappa}^{s_1 \downarrow}_{ \uparrow} \bar{\kappa}^{ \downarrow * }_{ \downarrow} =\bar{\kappa}^{ \uparrow}_{s_2 \downarrow} \bar{\kappa}^{s_1 \uparrow * }_{s_2 \uparrow}  = \bar{\kappa}^{s_1 \downarrow}_{s_2 \uparrow} \bar{\kappa}^{s_1 \uparrow * }_{s_2 \uparrow} = \frac{1}{4} \cos \phi \sin \phi, \\
 & \bar{\kappa}^{s_1 \uparrow}_{s_2 \downarrow} \bar{ \kappa}^{s_1 \uparrow * }_{ s_2 \downarrow}=\bar{\kappa}^{ s_1 \uparrow}_{s_2 \downarrow} \bar{\kappa}^{s_1 \downarrow * }_{s_2 \uparrow} = \bar{\kappa}^{ s_1 \downarrow}_{s_2 \uparrow} \bar{\kappa}^{s_1 \uparrow * }_{s_2 \downarrow} = \bar{\kappa}^{s_1 \downarrow}_{s_2 \uparrow} \bar{\kappa}^{s_1 \downarrow * }_{s_2 \uparrow} = \frac{1}{4} \sin^{2} \phi.
\end{aligned}
\end{equation*}
So, the eigenvalues of $\mathbb{R}$ are $\lbrace 1, 0, 0, 0 \rbrace$.  Thus 
\begin{equation}
\mathcal{C}_{s_{1_{a_{2}}} \mid s_{2_{b_{2}}}}\left( \rho_{s_{1_{a_{2}}}  s_{2_{b_{2}}}}\right) =1.
\end{equation}
Similar calculations follows that, $\mathcal{C}_{s_{1_{a_{2}}} \mid s_{2_{b_{1}}}}\left( \rho_{s_{1_{a_{2}}}  s_{2_{b_{1}}}}\right)=1$.

Likewise, we can also calculate the log-negativity~\cite{Zyczkowski98,Vidal02} for the density matrix $\rho_{s_{1_{a_{2}}} s_{2_{b_{2}}}}$ in Eq.~\eqref{Rho_s1a2_s2b2}. For that, we need the eigenvalues of the density matrix $\rho_{s_{1_{a_{2}}} }$  after taking the partial transpose with respect to $s_{2_{b_{2}}}$. The eigenvalues are found to be $\lbrace -\frac{1}{2},\frac{1}{2}, \frac{1}{2}, \frac{1}{2} \rbrace$. Thus the value of negativity is $\frac{1}{2}$ and so the log-negativity is given by 
\begin{equation}
E_{\mathcal{N}}\left( \rho_{s_{1_{a_{2}}}  s_{2_{b_{2}}}}\right) =1.
\end{equation}
Similar calculations give $E_{\mathcal{N}}\left(\rho_{s_{1_{a_{2}}}  s_{2_{b_{1}}}}\right) =1.$

All other monogamous measures of entanglement for qubit systems such as entanglement of formation~\cite{Oliveira14,Bai14PRL,Bai14PRA,Gou19}, log-negativity~\cite{Zyczkowski98,Vidal02}, Tsallis-q entropy~\cite{Kim10,Luo16}, R\'{e}nyi-$\alpha$ entanglement~\cite{Kim_Sanders10,Song16}, Unified-(q, s) entropy~\cite{Kim11,Khan19}, one-way distillable entanglement~\cite{Devetak05}, squashed entanglement~\cite{Christandl04,Brandao11} etc.~\cite{Plenio07} are calculable from the reduced density matrix. If one starts with the same reduced density matrix as in Eq.~\eqref{Rho_s1a2_s2b2}, one can easily show that all the above measures attain their respective maximum value for both the subsystems  $ \lbrace s_{1_{a_{2}}}, s_{2_{b_{2}}} \rbrace$ and $ \lbrace s_{1_{a_{2}}}, s_{2_{b_{1}}} \rbrace$ simultaneously, thereby violating the MoE. Thus, for any bipartite monogamous entanglement measure $\mathbb{E}$, we get
\begin{equation}
\mathbb{E}_{s_{1_{a_{2}}} \mid s_{2_{b_{2}}}}\left( \rho_{s_{1_{a_{2}}}  s_{2_{b_{2}}}}\right) = \mathbb{E}_{s_{1_{a_{2}}} \mid s_{2_{b_{1}}}}\left( \rho_{s_{1_{a_{2}}}  s_{2_{b_{1}}}}\right) = 1.
\end{equation}

This concludes the proof of Theorem 1.

\end{document}